\documentclass[runningheads]{llncs}
\usepackage{graphicx}
\usepackage{amssymb}
\usepackage{xcolor}
\usepackage{color, colortbl}
\usepackage{amsmath}
\usepackage{bbm}
\usepackage{multirow}
\usepackage{float}
\usepackage{arydshln}
\usepackage{pifont}
\newcommand{\cmark}{\ding{51}}%
\newcommand{\xmark}{\ding{55}}%
\usepackage{booktabs}
\usepackage{array}
\definecolor{Gray}{gray}{0.8}
\usepackage[width=122mm,left=12mm,paperwidth=146mm,height=193mm,top=12mm,paperheight=217mm]{geometry}

\graphicspath{{figs/}}

\newcommand{\ie}{\textit{i}.\textit{e}.\ }
\newcommand{\eg}{\textit{e}.\textit{g}.\ }
 
\newcommand*\samethanks[1][\value{footnote}]{\footnotemark[#1]}
\begin{document}
\title{Content-Preserving Unpaired Translation\\ from Simulated to Realistic Ultrasound Images}
\author{Devavrat Tomar\inst{1}\thanks{Both authors contributed equally.} \and 
Lin Zhang\inst{1}\samethanks \and
Tiziano Portenier\inst{1} \and
Orcun Goksel\inst{1,2}}
\authorrunning{Tomar et al.}
\titlerunning{Content-Preserving Unpaired Translation}
\institute{Computer-assisted Applications in Medicine, ETH Zurich, Switzerland
\and
Department of Information Technology, Uppsala University, Sweden
}
\maketitle              

\begin{abstract}
Interactive simulation of ultrasound imaging greatly facilitates sonography training.
Although ray-tracing based methods have shown promising results, obtaining realistic images requires substantial modeling effort and manual parameter tuning. In addition, current techniques still result in a significant appearance gap between simulated images and real clinical scans.
Herein we introduce a novel \textit{content-preserving} image translation framework (ConPres) to bridge this appearance gap, while maintaining the simulated anatomical layout.
We achieve this goal by leveraging both simulated images with semantic segmentations and unpaired in-vivo ultrasound scans.
Our framework is based on recent contrastive unpaired translation techniques and we propose a regularization approach by learning an auxiliary segmentation-to-real image translation task, which encourages the disentanglement of content and style.
In addition, we extend the generator to be class-conditional, which enables the incorporation of additional losses, in particular a cyclic consistency loss, to further improve the translation quality.
Qualitative and quantitative comparisons against state-of-the-art unpaired translation methods demonstrate the superiority of our proposed framework.

\end{abstract}
\section{Introduction}
Ultrasound (US) is a commonly used medical imaging modality that supports real-time and safe clinical diagnosis, in particular in gynecology and obstetrics.
However, the limited image quality and the hand-eye coordination required for probe manipulation necessitate extensive training of sonographers in image interpretation and navigation.
Volunteer access and realism of phantoms being limited for training, especially of rare diseases, computational methods become essential as simulation-based training tools.
To that end, interpolation of pre-acquired US volumes~\cite{goksel2009b} provide only limited image diversity.
Nevertheless, ray-tracing based methods have been demonstrated to successfully simulate images with realistic view-dependent ultrasonic artifacts, \eg refraction and reflection~\cite{burger2013real}.
Monte-Carlo ray-tracing~\cite{mattausch2018realistic} has further enabled realistic soft shadows and fuzzy reflections, while animated models and fusion of partial-frame simulations were also presented~\cite{starkov2019fusion}.
However, the simulation realism depends highly on the underlying anatomical models and the parametrization of tissue properties.
Especially the noisy appearance of ultrasound images with typical speckle patterns are nontrivial to parameterize.
Despite several approaches proposed to that end~\cite{mattausch2017image,starkov2019ultrasound,zhang2020scat}, images simulated from anatomical models  still lack realism, with the generated images appearing synthetic compared to real US scans.

Learning-based image translation techniques have received increasing interest in solving ultrasound imaging tasks, \eg cross-modality translation~\cite{jiao2020self}, image enhancement~\cite{jafari2020cardiac,zhang2020deep,zhang2021learning}, and semantic image synthesis~\cite{bargsten2020specklegan,tom2018simulating}.
The aim of these methods is to map images from a source domain to target domain, e.g. mapping low- to high-quality images.
Generative adversarial networks (GANs)~\cite{goodfellow2014generative} have been widely used in image translation due to their superior performance in generating realistic images compared to supervised losses.
In the paired setting, where images in the source domain have a corresponding ground truth image in the target domain, a combination of supervised per-pixel losses and a conditional GAN loss~\cite{mirza2014conditional} has shown great success on various translation tasks~\cite{isola2017image}.
In the absence of paired training samples, the translation problem becomes under-constrained and additional constraints are required to learn a successful translation. 
To tackle this issue, a cyclic consistency loss (cycleGAN) was proposed~\cite{zhu2017unpaired}, where an inverse mapping from target to source domain is learned simultaneously, while a cycle consistency is ensured by minimizing a reconstruction loss between the output of the inverse mapping and the source image itself.
Recent works have extended and applied cycle consistency on multi-domain translation~\cite{almahairi2018augmented,choi2018stargan,zhu2017toward}.
Cycle consistency assumes a strong bijective relation between the domains.
To relax the bijectivity assumption and reduce the training burden, Park et al.~\cite{park2020contrastive} proposed an alternative with a single-sided unpaired translation technique with contrastive learning.
For US simulation, the standard cycleGAN was used in~\cite{vitale2019improving} to improve the realism of simulated US image frames, however, this method is prone to generate unrealistic deformations and hallucinated features.

In this work, we aim to improve the realism of computationally-simulated US images by converting their appearance to that of real in-vivo US scans, while preserving their anatomical content and view-dependent artefacts originating from the preceeding computational simulation.
We build our framework on a recent contrastive unpaired translation framework~\cite{park2020contrastive} and introduce several contributions to improve translation quality.
In particular, to encourage content preservation, we propose to (i)~constrain the generator with the accompanying semantic labels of simulated images by learning an auxiliary segmentation-to-real image translation task; and (ii) apply a class-conditional generator, which in turn enables the incorporation of a cyclic loss.

\section{Method}
\begin{figure}[t]
\centering
\includegraphics[width=\textwidth]{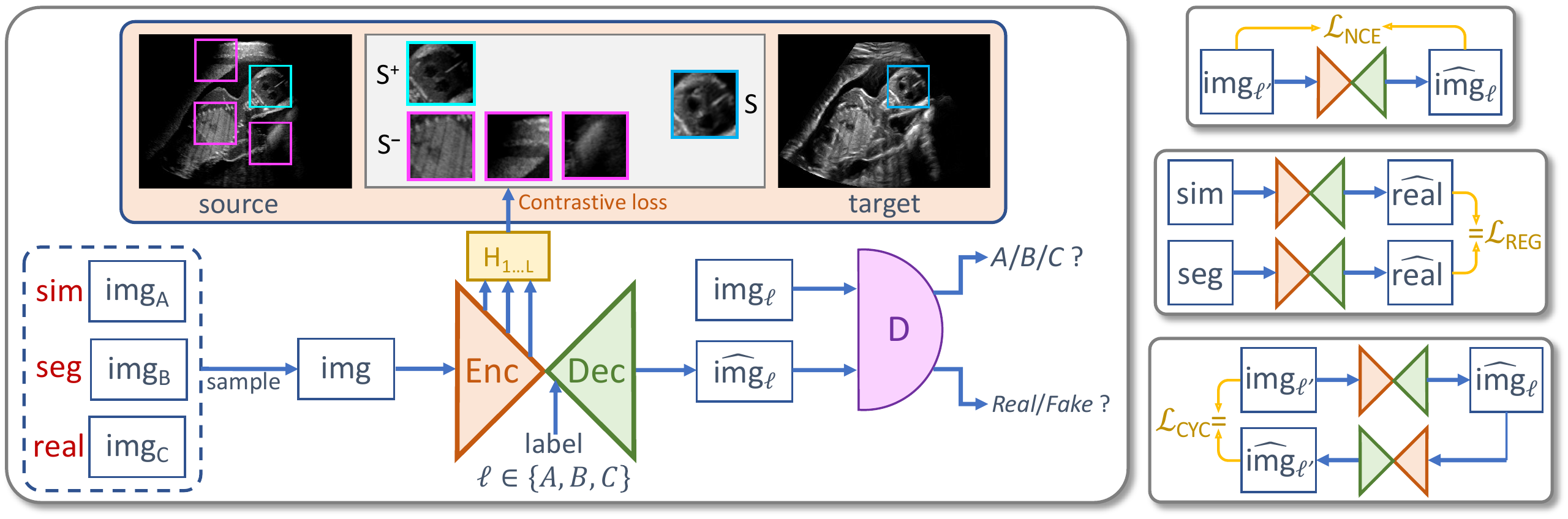}
\caption{(Left) Overview of our proposed framework. (Right) Illustrations of some of the loss functions used to train our model.}
\label{fig:nn}
\end{figure}

Given unpaired source images $X=\{x\in\mathbb{X}\}$ and target images $Y=\{y\in\mathbb{Y}\}$, we aim to learn a generator function $G:\mathbb{X}\mapsto\mathbb{Y}$, such that mapped images $G(x)$ have similar appearance (style) as images in $Y$, while preserving the structural content of the input image $x$.
To achieve this goal, we divide $G$ into an encoder $G_\mathrm{enc}$ and a decoder $G_\mathrm{dec}$. $G_\mathrm{enc}$ is restricted to extract content-related features only, while $G_\mathrm{dec}$ learns to generate a desired target appearance using a patch contrastive loss. Combined with both cyclic and semantic regularizations, we design a multi-domain translation framework consisting of a single generator and discriminator (Fig.~\ref{fig:nn}).

\vspace{1ex}\noindent{\bf Adversarial loss. }
We adopt the patchGAN discriminator~\cite{park2020contrastive} that discriminates real and fake images using a least squares GAN loss:
\begin{equation}
    \mathcal{L}_{\text{GAN}}(X,Y)=\mathbb{E}_y \log [(D(y)-1)^2] + \mathbb{E}_{x} \log [D(G(x))^2]\ .
\end{equation}

\vspace{1ex}\noindent{\bf Contrastive loss. }
An unpaired contrastive translation framework (CUT) is presented in~\cite{park2020contrastive} that maximizes mutual information between image patches in the source and target domain to maintain the content of source images.
The core of this approach is to enforce each translated patch to be ($i$)~similar to the corresponding input patch, while ($ii$)~different from any other input patches.
For the similarity assessment, image patches are represented by hidden features of $G_\mathrm{enc}$.
A multi-layer perceptron (MLP) $H_l$ with two hidden layers is then used to map the chosen encoder features $h_l$ to an embedded representation $z_l=H_l(h_l) \in \mathbb{R}^{S_l \times C_l}$ with $S_l$ spatial locations and $C_l$ channels, where $h_l=G_\mathrm{enc}^l(x)$ is the $l$-th hidden layer of $G_\mathrm{enc}$.
For each spatial location $s$ in $z_l$, the corresponding patch feature vector $z_l^{s+} \in \mathbb{R}^{C_l}$ is then the positive sample and the features at any other locations are the negatives $z_l^{s-} \in \mathbb{R}^{(S_l-1)\times C_l}$.
The corresponding patch feature $\hat{z_l}^s = h_l( G_{enc}^l(\hat{y})) \in \mathbb{R}^{C_l}$ of the output image $\hat{y}$ acts as the query.
The contrastive loss is defined as the cross-entropy loss
\begin{align}
     l(\hat{z}_l^s, z_l^{s+}, z_l^{s-}) = - \log \left[ \frac{\exp(\hat{z}_l^s\cdot z_l^{s+}/\tau)}{\exp(\hat{z}_l^s\cdot z_l^{s+}/\tau)+\sum_{k=1}^{S_l-1}\exp(\hat{z}_l^s\cdot z_{l,k}^{s-}/\tau)} \right],
\end{align}
with the temperature parameter $\tau$ set to 0.07, following~\cite{park2020contrastive}.
Using features from multiple encoder depths allows us to enforce patch similarity on multiple scales, leading to the following noise contrastive estimation (NCE) loss
\begin{equation}
    \mathcal{L}_\text{NCE}(X)=\mathbb{E}_{x}\sum_{l=1}^{L}\sum_{s=1}^{S_l} l(\hat{z}_l^s, z_l^{s+}, z_l^{s-}),
\end{equation}
where $L$ is the number of layers used for computing the loss.
To encourage the generator to translate the domain-specific image appearance only,  $\mathcal{L}_{\text{NCE}}$ is also evaluated on the target domain $\mathbb{Y}$, which acts as an identity loss, similarly to the cyclic consistency loss in~\cite{zhu2017unpaired}.
The final objective in CUT~\cite{park2020contrastive} is defined as
\begin{equation}
    \mathcal{L}_\text{CUT}(X,Y) = \mathcal{L}_\text{GAN}(X,Y) + \mathcal{L}_\text{NCE}(X) + \mathcal{L}_\text{NCE}(Y).
\end{equation}

\vspace{1ex}\noindent{\bf Semantic-consistent regularization. }
To encourage the disentanglement of content and style, we leverage available surrogate segmentation maps $S = \{s\in\mathbb{S}\}$ of the simulated images (sim).
In addition to sim-to-real translation, our generator then learns to also synthesize real images from segmentation maps (seg), \ie seg-to-real translation.
Since segmentation maps contain only content and no style, it is ensured that, after passing $G_\text{enc}$, there is no style left in the features, therefore $G_\text{dec}$ has to introduce styles entirely from scratch. 
Learning this auxiliary task thus helps to prevent style leakage from $G_\text{enc}$, enforcing $G_\text{enc}$ to extract only content-relevant features.
In this modified CUT framework with semantic input (CUT\texttt{+}S), we minimize
\begin{equation}
    \mathcal{L}_\text{CUT\texttt{+}S} = \mathcal{L}_\text{CUT}(X,Y) +
    \mathcal{L}_\text{GAN}(S,Y) +
    \mathcal{L}_\text{NCE}(S)\,.
\end{equation}
In addition, we regularize $G$ to generate the same output for paired seg and sim, thus explicitly incorporating the semantic information of simulated images into the generator. We achieve this by minimizing the following semantic-consistent regularization loss: $\mathcal{L}_\text{REG}(X,S)=\mathbb{E}_{x,s}||G(x)-G(s)||_1$. 
Our consistency-based training objective then becomes:
\begin{equation}
    \mathcal{L}_\text{CUT\texttt{+}SC} = \mathcal{L}_\text{CUT\texttt{+}S} +
    \lambda_\text{REG} \mathcal{L}_\text{REG}(X,S)\,.
\end{equation}

\vspace{1ex}\noindent{\bf Multi-domain translation. }
In preliminary experiments, we observed that despite the identity contrastive loss and semantic inputs, the generator still alters the image content, since the above losses do not explicitly enforce the structural consistency between input and translated images.
To mitigate this issue, we require a cyclic consistency loss similar to~\cite{zhu2017unpaired}.
For this purpose, we extend the so-far single-direction translation to a multi-domain translation framework, while keeping a unified (now conditional) generator and discriminator, inspired by StarGAN~\cite{choi2018stargan}.
Here, $G_\text{dec}$ is trained to transfer the target appearance, conditioned by the target class label $\ell\in\{\mathbb{A},\mathbb{B},\mathbb{S}\}$ given the classes $\mathbb{A}$ simulated image, $\mathbb{B}$ real image, and $\mathbb{S}$ semantic map.
The class label is encoded as a one-hot vector and concatenated to the input of the decoder.
The cyclic consistency loss is then defined as
\begin{equation}
    \mathcal{L}_\text{CYC}(X)=\mathbb{E}_{x,\ell,\ell'}||x-G(G(x,\ell),\ell')||_1,
\end{equation}
where $\ell'$ is the class label of the input image and $\ell$ is label of the target class.

\vspace{1ex}\noindent{\bf Classification loss. }
To enable class-dependent classification (CLS) with the discriminator~\cite{choi2018stargan}, $D$ tries to predict the correct domain class label $\ell'$ for a given \emph{real} image $x$ as an auxiliary task, \ie
\begin{equation}
    \mathcal{L}_\text{CLS,r} (X) = \mathbb{E}_{x,\ell'}[-\log D(\ell'|x)],
\end{equation}
while $G$ tries to fool $D$ with \emph{fake} images to be classified as target domain $\ell$ by minimizing
\begin{equation}
    \mathcal{L}_\text{CLS,f} (X)  = \mathbb{E}_{x,\ell}[-\log D(\ell|G(x,\ell))].
\end{equation}

\vspace{1ex}\noindent{\bf Final objective }
For our final model (ConPres), the training objective is evaluated by randomly sampling two pairs of domains $(X_i,Y_i)\in \{ (\mathbb{A},\mathbb{B},\mathbb{S})  \backslash X_i\neq Y_i\}$ for $i=[1,2]$, given the following discriminator and generator losses 
\begin{align}
    \mathcal{L}^\text{D}_\text{ConPres} &= \textstyle\sum_{i=1}^2 
    -\mathcal{L}_\text{GAN}(X_i, Y_i) + \lambda_\text{CLS,r} \mathcal{L}_\text{CLS,r} (X_i), \\
    \mathcal{L}^\text{G}_\text{ConPres} &= \textstyle\sum_{i=1}^2 
    \mathcal{L}_\text{CUT}(X_i,Y_i)+
    \lambda_\text{CLS,f} \mathcal{L}_\text{CLS,f} (X_i)
    + \lambda_\text{CYC} \mathcal{L}_\text{CYC} (X_i) \nonumber
    \\[-0.5ex] & \qquad \qquad + \mathbbm{1}_{[\,(X_1=\mathbb{A}\land X_2=\mathbb{S})\,\lor\,(X_1=\mathbb{S}\land X_2=\mathbb{A})\,]} \lambda_{\text{REG}} \mathcal{L}_{\text{REG}}(X_1, X_2)
\end{align}
with the indicator function $\mathbbm{1}_{[.]}$ and the hyperparameters $\lambda_{\{\cdot\}}$ for weighting loss components.
We set $\lambda_{\text{REG}}$$=$$0$ when the two source domains are not $\mathbb{A}$ and $\mathbb{S}$.

\section{Experiments and Results}

\begin{figure}[t]
\centering
\includegraphics[width=0.24\textwidth]{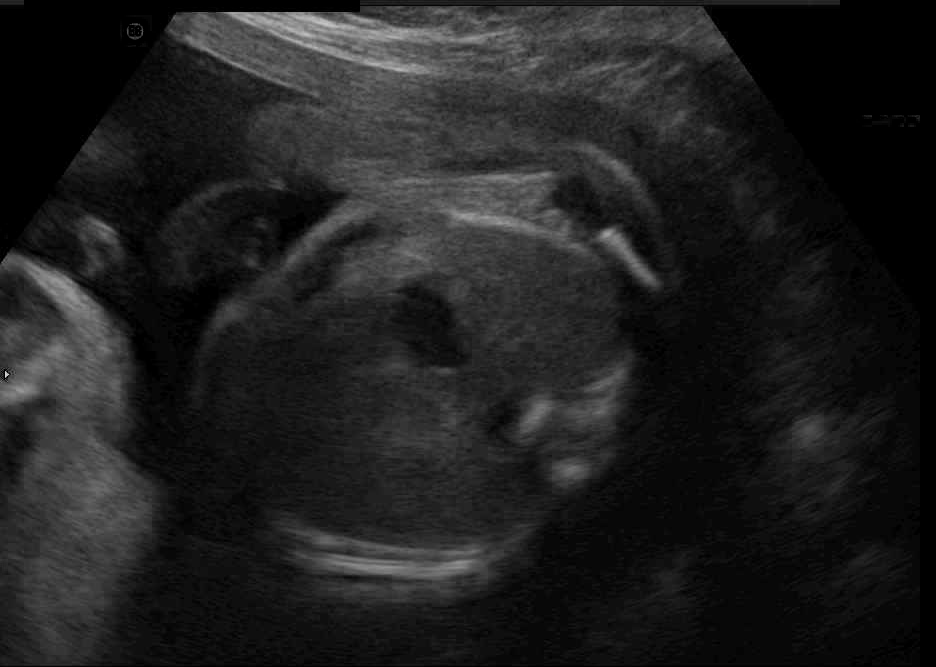}
\includegraphics[width=0.24\textwidth]{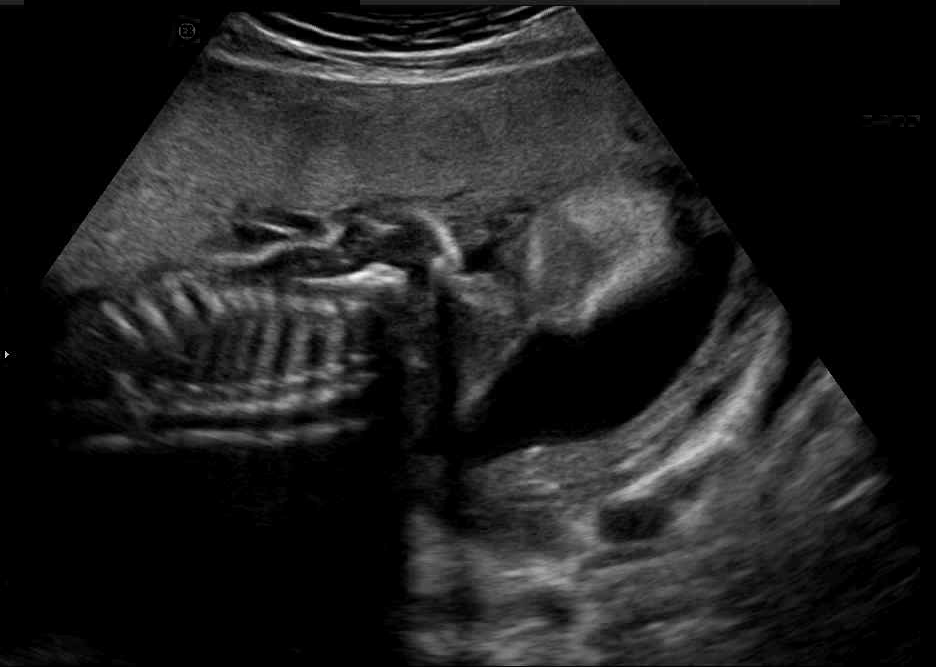}
\includegraphics[width=0.24\textwidth]{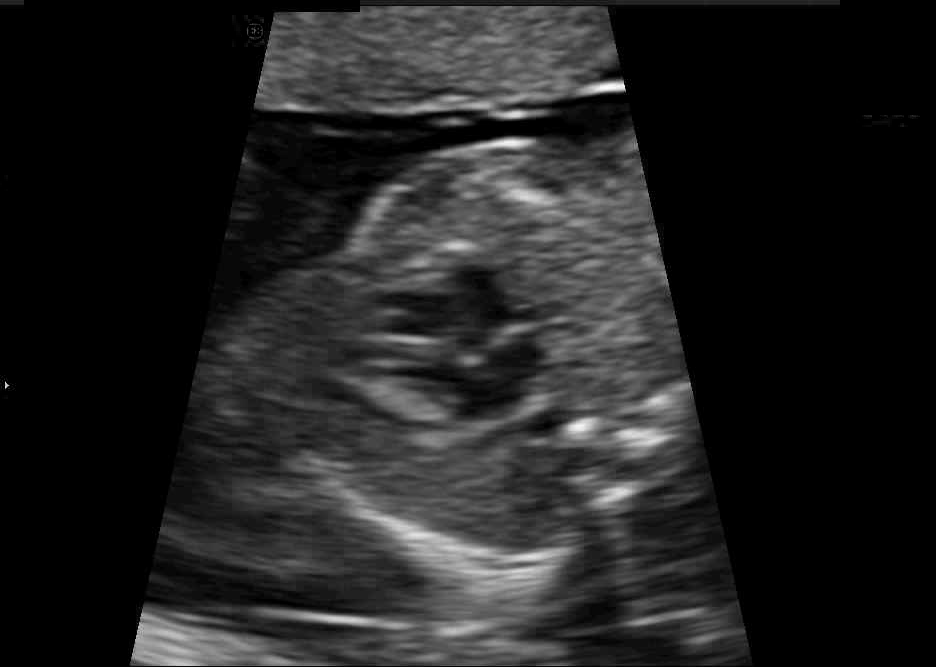}
\includegraphics[width=0.24\textwidth]{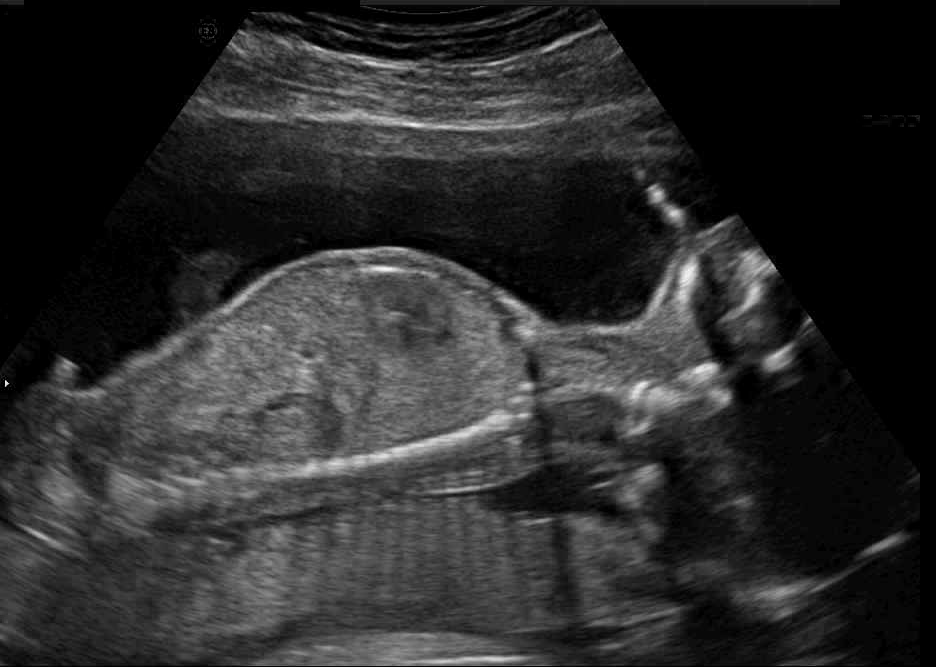}
\caption{Examples of in-vivo images used to train our model.}
\label{fig:real_exp}
\end{figure}

\vspace{1ex}\noindent{\bf Real in-vivo images. }
22 ultrasound sequences were collected using a GE Voluson E8 machine during standard fetal screening exams of 8 patients. Each sequence is several seconds long.
We extracted all 4427 frames and resize them to $256\times 354$, see Fig.~\ref{fig:real_exp} for some examples. 
The resulting image set was randomly split into training-validation-test sets by a 80-10-10\% ratio.

\begin{figure}[t]
\centering
\resizebox{0.89\textwidth}{!}{%
\begin{tabular}{@{}lccccc@{}}
\rotatebox{90}{\hspace{.3em}Simulation}    & \includegraphics[width=.20\textwidth]{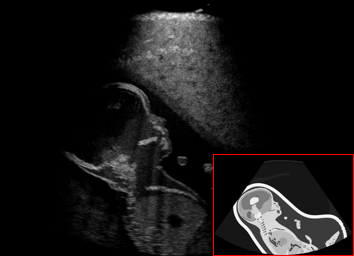} & 
\includegraphics[width=.20\textwidth]{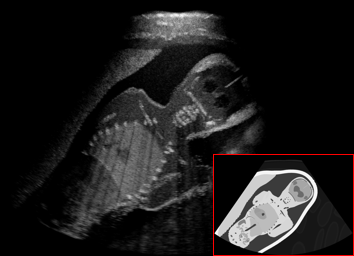} & \includegraphics[width=.20\textwidth]{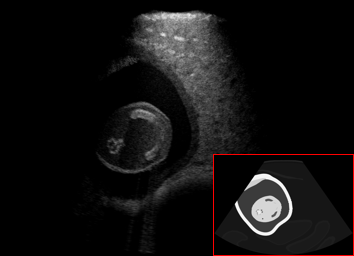} & \includegraphics[width=.20\textwidth]{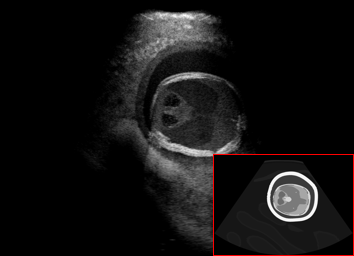} &
\includegraphics[width=.20\textwidth]{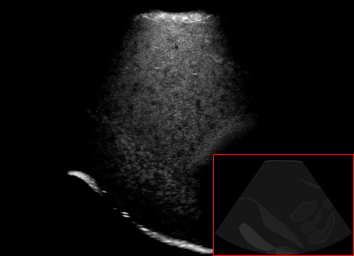}\\
\rotatebox{90}{\hspace{0.2em}CycleGAN} & \includegraphics[width=.20\textwidth]{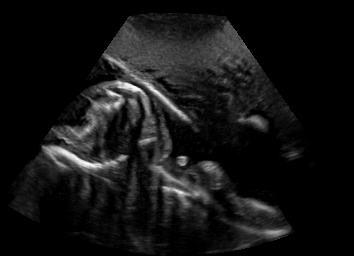} & 
\includegraphics[width=.20\textwidth]{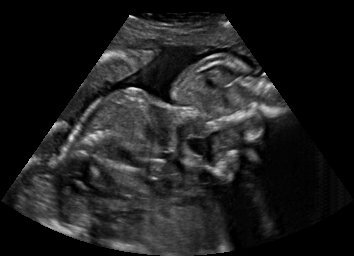} & \includegraphics[width=.20\textwidth]{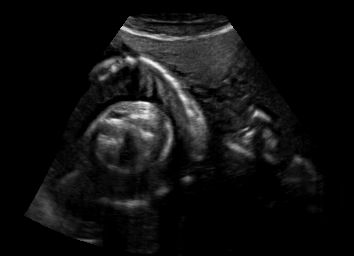} & \includegraphics[width=.20\textwidth]{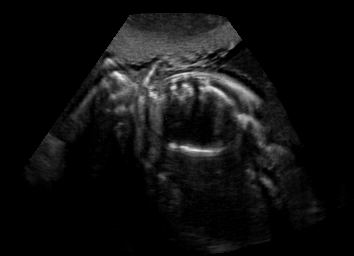} &
\includegraphics[width=.20\textwidth]{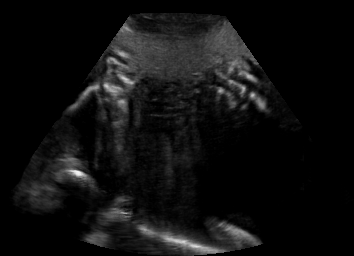}\\
\rotatebox{90}{\hspace{1em}SASAN} & \includegraphics[width=.20\textwidth]{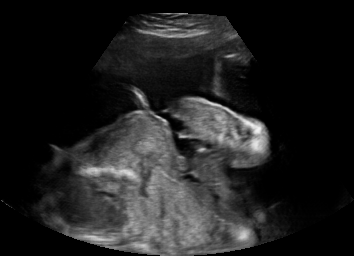} & 
\includegraphics[width=.20\textwidth]{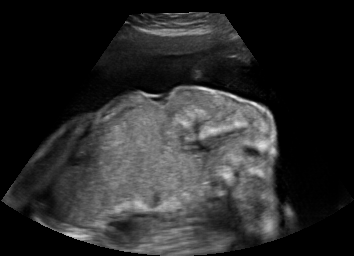} & \includegraphics[width=.20\textwidth]{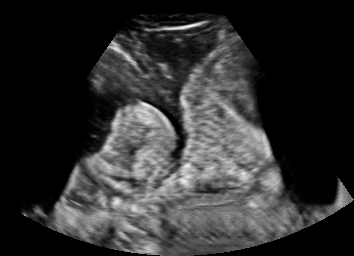} & \includegraphics[width=.20\textwidth]{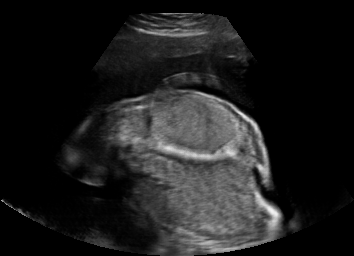} &
\includegraphics[width=.20\textwidth]{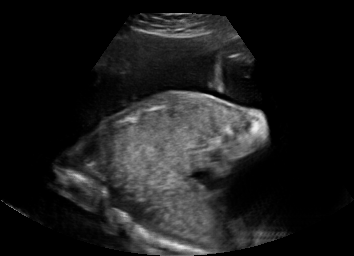}\\
\rotatebox{90}{\hspace{1.5em}CUT}      & \includegraphics[width=.20\textwidth]{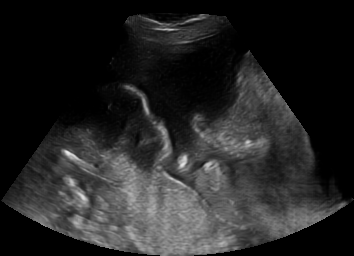} & 
\includegraphics[width=.20\textwidth]{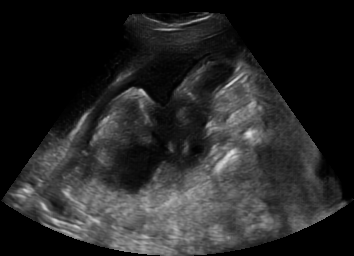} 
& \includegraphics[width=.20\textwidth]{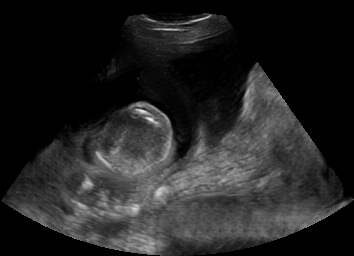} & \includegraphics[width=.20\textwidth]{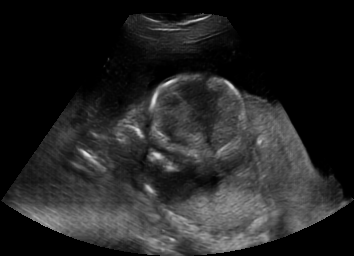} &
\includegraphics[width=.20\textwidth]{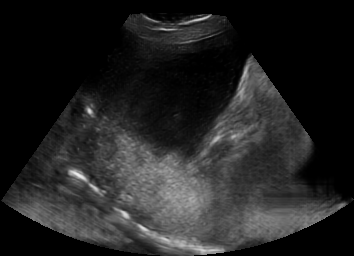}\\
\rotatebox{90}{\hspace{0.5em}StarGAN}  & \includegraphics[width=.20\textwidth]{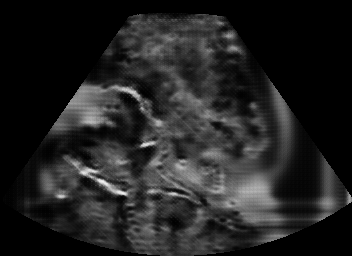} & 
\includegraphics[width=.20\textwidth]{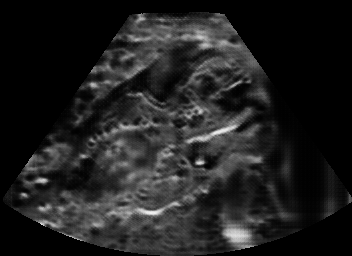} & \includegraphics[width=.20\textwidth]{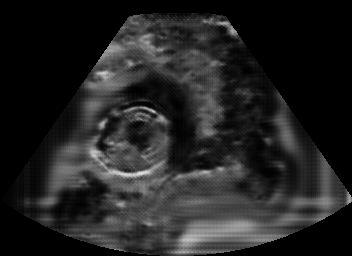} & \includegraphics[width=.20\textwidth]{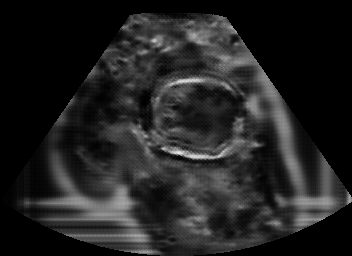} &
\includegraphics[width=.20\textwidth]{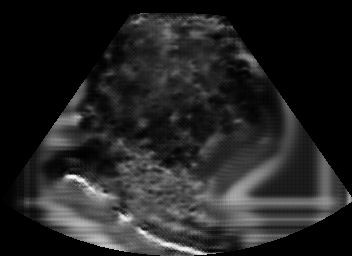}\\
\rotatebox{90}{\hspace{1em}CUT\texttt{+}S}     & \includegraphics[width=.20\textwidth]{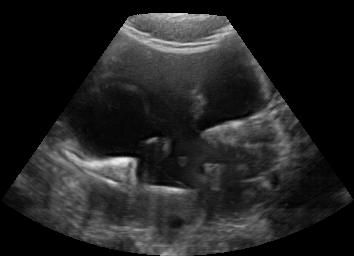} & 
\includegraphics[width=.20\textwidth]{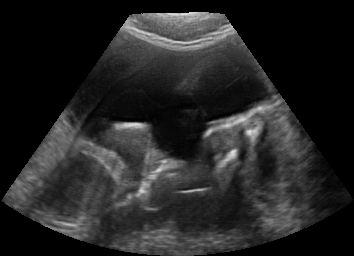} & 
\includegraphics[width=.20\textwidth]{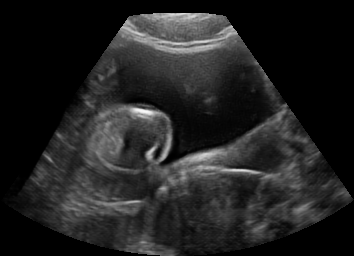} & \includegraphics[width=.20\textwidth]{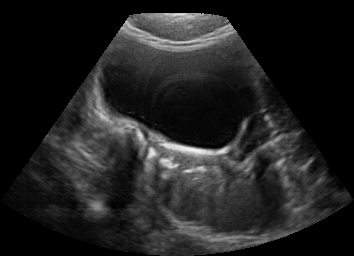} &
\includegraphics[width=.20\textwidth]{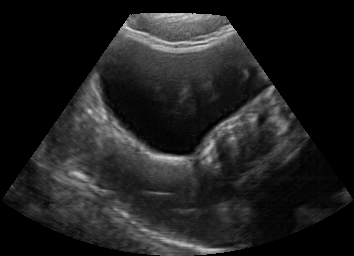}\\
\rotatebox{90}{\hspace{0.7em}CUT\texttt{+}SC}   & \includegraphics[width=.20\textwidth]{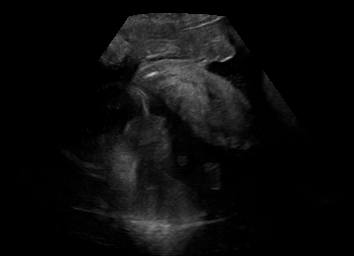} & 
\includegraphics[width=.20\textwidth]{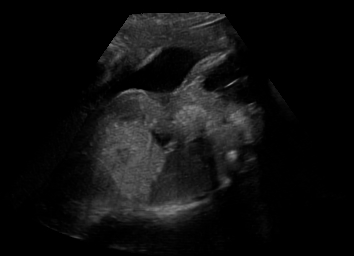} & \includegraphics[width=.20\textwidth]{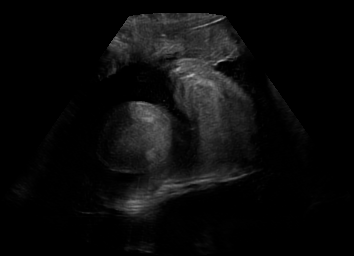} & \includegraphics[width=.20\textwidth]{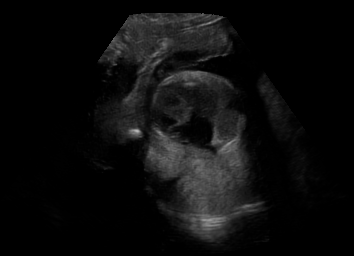} &
\includegraphics[width=.20\textwidth]{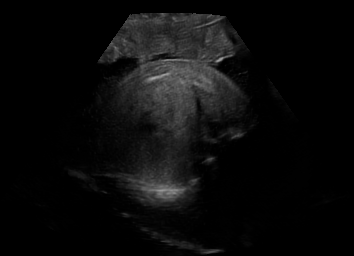} \\
\rotatebox{90}{\hspace{0.5em}ConPres}  & \includegraphics[width=.20\textwidth]{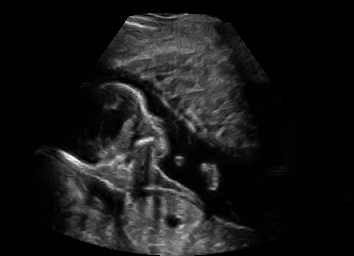} & 
\includegraphics[width=.20\textwidth]{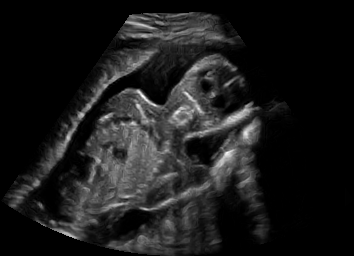} & 
\includegraphics[width=.20\textwidth]{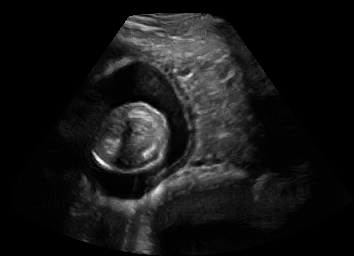} & \includegraphics[width=.20\textwidth]{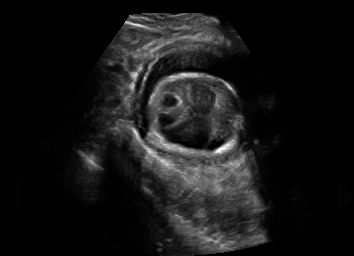} &
\includegraphics[width=.20\textwidth]{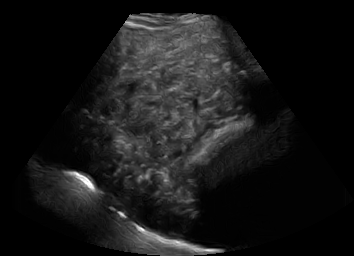}
\end{tabular}%
}
\caption{Qualitative results,
with images masked by foreground in segmentations.}
\label{fig:qualitative_results}
\end{figure}

\vspace{1ex}\noindent{\bf US simulation. }
We used a ray-tracing framework to render B-mode images from a geometric fetal model, by simulating a convex probe placed at multiple locations and orientations on the abdominal surface, with imaging settings listed in Tab~\ref{tab:sim_params}. 
At each location, simply rasterizing a cross-section through the triangulated anatomical surfaces at the ultrasound center imaging plane provided corresponding semantic maps. 
Fig.~\ref{fig:qualitative_results} shows example B-mode images with corresponding semantic maps.
A total of $6669$ simulated frames were resized to $256\times 354$ and randomly split into training-validation-test sets by 80-10-10\%.
\begin{table}
\caption{Anatomical model and imaging parameters for the US simulation}
\label{tab:sim_params}
\centering
\begin{tabular}{l||r}
\rowcolor[HTML]{C0C0C0} Parameter & Value \\
\hline 
Triangles fetus & 400k\\
Triangles mother & 275k\\
Image depth & 15.0 cm \\
Elevational layer & 3
\end{tabular}\hfil
\begin{tabular}{l||r}
\rowcolor[HTML]{C0C0C0} Parameter & Value \\
\hline 
Transducer frequency & 8 MHz\\
Field-of-view & 70$^{\circ}$\\
Axial samples & 3072 \\
Ray/scanline & 32
\end{tabular}
\end{table}

\vspace{1ex}\noindent{\bf Metrics. }
We use the following metrics to quantitatively evaluate our method:
\begin{itemize}
\item[$\bullet$] {\bf Structural similarity index} (SSIM) measures the structural similarity between simulated and translated images, quantifying content preservation.
We evaluate SSIM within regions having content in simulated images.
\item[$\bullet$] {\bf Fr\'{e}chet inception distance} (FID)~\cite{heusel2017gans} measures the feature distribution difference between two sets of images, herein real and translated, using feature vectors of Inception network.
Since a large number of samples is required to reduce estimation bias, we use the \emph{pre-aux} layer features, which has a smaller dimensionality than the default pooling layer features.
\item[$\bullet$] {\bf Kernel inception distance }(KID)~\cite{binkowski2018demystifying} is an alternative unbiased metric to evaluate GAN performance.
KID is computed as the squared maximum mean-discrepancy between the features of Inception network.
We use the default pooling layer features of Inception, to compute this score.
\end{itemize}

\vspace{1ex}\noindent{\bf Implementation details. }
We use a least-squares GAN loss with patchGAN discriminator as in~\cite{choi2018stargan}. 
The generator follows an encoder-decoder architecture, where the encoder consists of two stride-2 convolution layers followed by 4 residual blocks, while the decoder consists of 4 residual blocks followed by two convolution layers with bilinear upsampling. 
For architectural details, please see the supplementary material.
To compute the contrastive loss, we extract features from the input layer, the stride-2 convolution layers, and the outputs of the first three residual blocks of the encoder.
For CUT and its variants CUT\texttt{+}S and CUT\texttt{+}SC, we used the default layers in~\cite{park2020contrastive}.
To compute $\lambda_{\text{REG}}$, the sampled simulated and segmentation images in each batch are paired.
We used Adam~\cite{kingma2014adam} optimizer to train our model for 100 epochs with an $l_2$ regularization of $10^{-4}$ on model parameters with gradient clipping and $\beta=(0.5, 0.999)$. 
We set $\lambda_{\text{CLS},*}= $ 0.1, $\lambda_{\text{REG}}=1$ and $\lambda_{\text{CYC}}=10$.
We set the hyper-parameters based on similar losses in the compared implementations, for comparability; while we grid-searched the others, \eg $\lambda_{\text{REG}}$, for stable GAN training.
We implemented our model in PyTorch~\cite{paszke2019pytorch}.
For KID and FID computations, we used the implementation of~\cite{obukhov2020torchfidelity}.

\vspace{1ex}\noindent{\bf Comparative study. }
We compare our proposed ConPres to several state-of-the-art unpaired image translation methods:
\vspace{-1ex}\begin{itemize}
\item[$\bullet$]{\bf CycleGAN}~\cite{zhu2017unpaired}: A conventional approach with cyclic consistency loss.
\item[$\bullet$]{\bf SASAN}~\cite{tomar2021self}: CycleGAN extension with self-attentive spatial adaptive normalization, leveraging semantic information to retain anatomical structures, while translating using spatial attention modules and SPADE layers~\cite{park2019semantic}.
\item[$\bullet$]{\bf CUT}~\cite{park2020contrastive}: Unpaired contrastive framework for image translation.
\item[$\bullet$]{\bf StarGAN}~\cite{choi2018stargan}: A unified GAN framework for multi-domain translation.
\end{itemize}\vspace{-1ex}
We used the official implementations and default hyperparameters for training all the baselines.
To assess the effectiveness of the proposed architecture and losses, we also compare with the models CUT\texttt{+}S (CUT plus the seg-to-real translation) and CUT\texttt{+}SC (CUT\texttt{+}S plus $\mathcal{L}_\text{REG}$).

In Fig.~\ref{fig:qualitative_results} we show that only learning an auxiliary seg-to-real translation, i.e.\ CUT\texttt{+}S, cannot guide the network to learn the semantics of simulated images.

CUT\texttt{+}SC with the loss term $\mathcal{L}_{\text{REG}}$ largely reduces hallucinated image content, although it still fails to generate fine anatomical details.
With the multi-domain conditional generator and additional losses of ConPres, translated images preserve content and feature a realistic appearance.
Training without $\mathcal{L}_{\text{NCE}}$ leads to training instability. 

\vspace{1ex}\noindent{\bf Comparison to state-of-the-art. }
As seen qualitatively from the examples in Fig.~\ref{fig:qualitative_results}, 
our method substantially outperforms the alternatives in terms of content preservation, while translating realistic US appearance.
CycleGAN, SASAN, and CUT hallucinate inexistent tissue regions fail to generate fine anatomical structures, \eg the ribs.
StarGAN fails to generate faithful ultrasound speckle appearance, which leads to highly unrealistic images.
Our method ConPres preserves anatomical structures, while enhancing the images with a realistic appearance.
It further faithfully preserves acoustic shadows, even without explicit enforcement.
However, as seen from the last column, the refraction artefact appears artificial in the images translated by all the methods.
Note that although the imaging field-of-view (FoV) and probe opening in the simulation is significantly different from the real in-vivo images~(Fig.~\ref{fig:real_exp}) used for training, our ConPres maintains the input FoV closely compared to previous state-of-the-art.
The results in Tab~\ref{tab:quant_results} quantitatively confirm the superiority of our method.
Note that SSIM and FID/KID are used to measure translation performance from two different and sometimes competing aspects, with the former metric for quantifying structure preservation and the latter metrics for image realism.

\newcommand{\std}[1]{{\scriptsize$\pm$#1}}
\begin{table}[t]
\setlength{\tabcolsep}{2pt}
\caption{Quantitative metrics and ranking from the user study (mean\std{std}). Best results are marked bold. "Seg" gives if semantic maps are used as network input.}
\centering
   \begin{tabular}{c|c|l||c|c|c||c} 
 &\multicolumn{1}{c|}{\cellcolor[HTML]{C0C0C0}Seg}
 &\multicolumn{1}{c||}{\cellcolor[HTML]{C0C0C0}Method}
 &\multicolumn{1}{c|}{\cellcolor[HTML]{C0C0C0}SSIM $\uparrow$}
 &\multicolumn{1}{c|}{\cellcolor[HTML]{C0C0C0}FID $\downarrow$}
 &\multicolumn{1}{c||}{\cellcolor[HTML]{C0C0C0}KID $\downarrow$}
 &\multicolumn{1}{c}{\cellcolor[HTML]{C0C0C0}Ranking $\in[1,6]$$\downarrow$}\\
\hline
&--- &Simulation & ---  &2.37 &0.41 &3.98\std{1.35} \\
\hdashline
\parbox[t]{2mm}{\multirow{4}{*}{\rotatebox[origin=c]{90}{Others}}}
&\xmark &CycleGAN~\cite{zhu2017unpaired} & 71.73\std{5.18}  &1.78 &0.32 &2.86\std{1.27} \\
&\cmark &SASAN~\cite{tomar2021self} & 68.20\std{4.00}  &2.36 &0.39 &3.59\std{1.55} \\
&\xmark &CUT~\cite{park2020contrastive} & 67.28\std{4.62}  &1.77 &0.31 & 2.92\std{1.20} \\
&\cmark &StarGAN~\cite{choi2018stargan} & 63.62\std{4.82}  &1.93 &0.47 &5.76\std{0.61} \\
\hdashline
\parbox[t]{2mm}{\multirow{3}{*}{\rotatebox[origin=c]{90}{Ours}}}
&\cmark &CUT\texttt{+}S  & 68.88\std{4.63}  &2.25 &0.41 & \\
&\cmark &CUT\texttt{+}SC  & \bf{80.56}\std{2.11}  &1.87 &0.38 & \\
&\cmark &ConPres  & 72.13\std{4.58}  &\bf{1.51} &\bf{0.24} &\bf{1.89}\std{1.07}
\end{tabular}
\label{tab:quant_results}
\end{table}
\begin{figure}[t]
\centering
\includegraphics[width=0.19\textwidth]{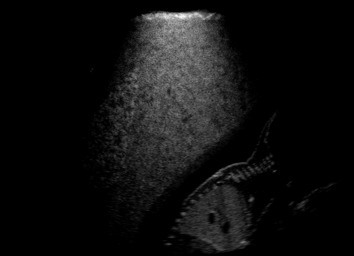}
\includegraphics[width=0.19\textwidth]{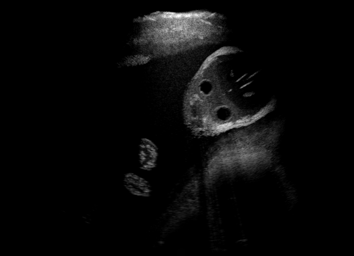}
\includegraphics[width=0.19\textwidth]{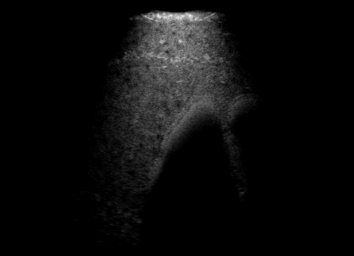}
\includegraphics[width=0.19\textwidth]{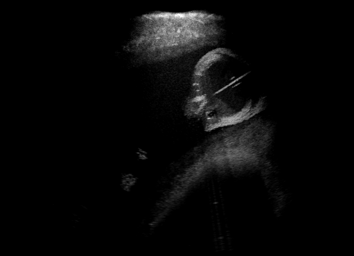}
\includegraphics[width=0.19\textwidth]{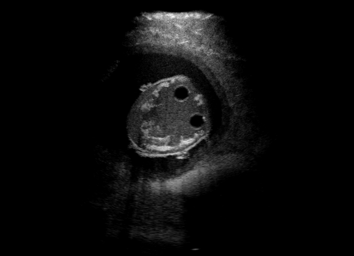}\\
\includegraphics[width=0.19\textwidth]{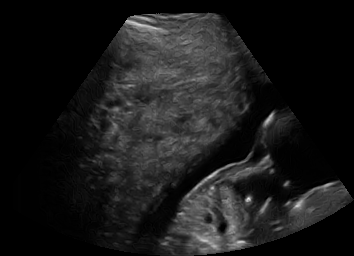}
\includegraphics[width=0.19\textwidth]{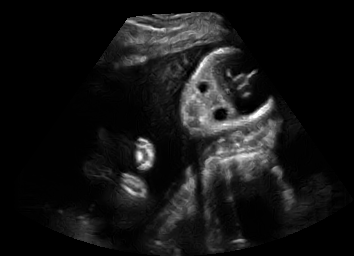}
\includegraphics[width=0.19\textwidth]{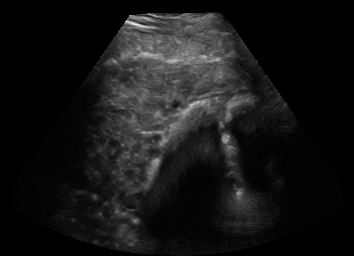}
\includegraphics[width=0.19\textwidth]{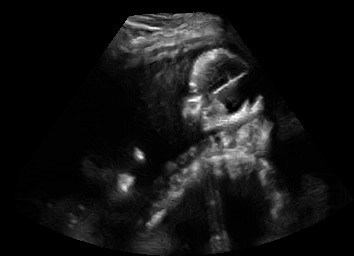}
\includegraphics[width=0.19\textwidth]{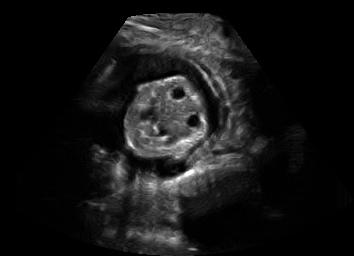}\\
\includegraphics[width=0.19\textwidth]{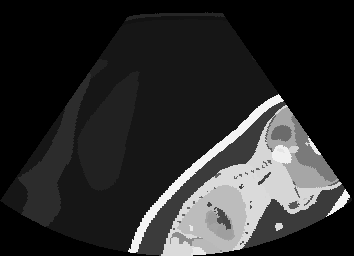}
\includegraphics[width=0.19\textwidth]{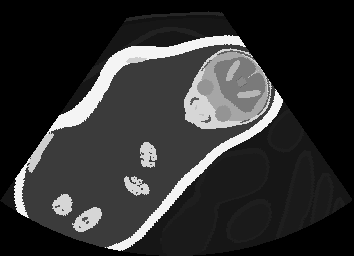}
\includegraphics[width=0.19\textwidth]{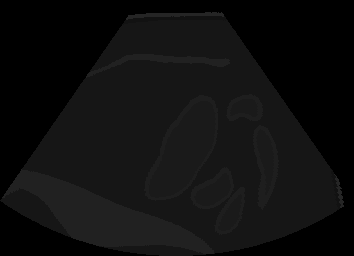}
\includegraphics[width=0.19\textwidth]{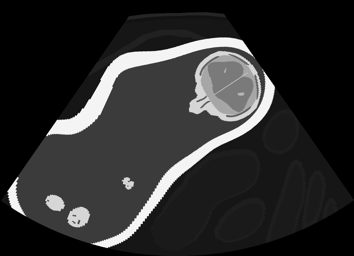}
\includegraphics[width=0.19\textwidth]{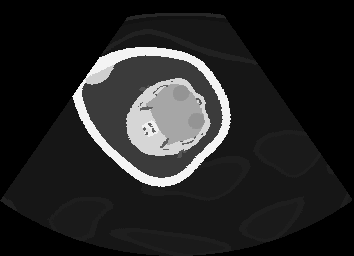}
\caption{Other sim-to-real image translation examples. Top: input simulated images, middle: translated real images, bottom: semantic maps for reference.}
\label{fig:add_results}
\end{figure}

\vspace{1ex}\noindent{\bf A user study }was performed with 18 participants (14 technical and 4 clinical ultrasound experts) to evaluate the realism of translated images for 20 US frames with example images shown in Figs.~\ref{fig:qualitative_results} and~\ref{fig:add_results}.
For each frame, a separate questionaire window opened in a web interface, presenting the participants with six candidate images including the input simulated famre and its translated versions using CUT, CycleGAN, SASAN, StarGAN, and ConPres.
As a reference for the given ultrasound machine appearance, we also showed a fixed set of 10 real in-vivo images.
The participants were asked to rank the candidate images based on ``their likelihood for being an image from this machine''.
The average rank score is reported in Tab~\ref{tab:quant_results}.
Based on a paired Wilcoxon signed rank test, our method is significantly superior to any competing method (all p-values $<10^{-18}$).

\begin{figure}[t]
\centering
\includegraphics[width=0.19\textwidth]{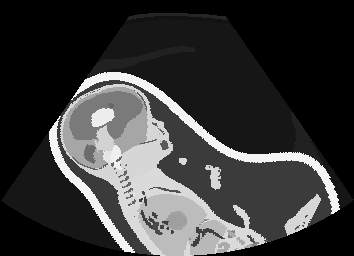}
\includegraphics[width=0.19\textwidth]{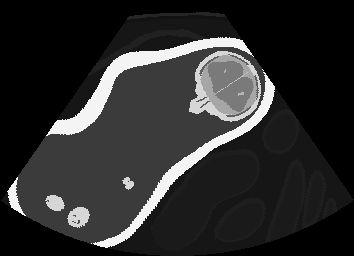}
\includegraphics[width=0.19\textwidth]{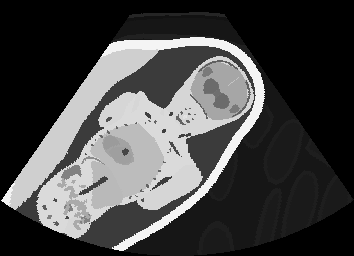}
\includegraphics[width=0.19\textwidth]{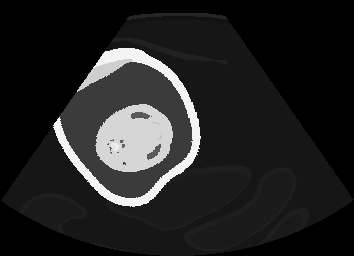}
\includegraphics[width=0.19\textwidth]{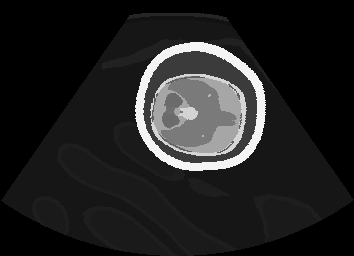}
\includegraphics[width=0.19\textwidth]{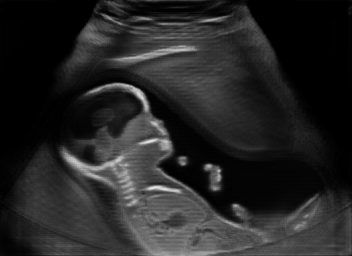}
\includegraphics[width=0.19\textwidth]{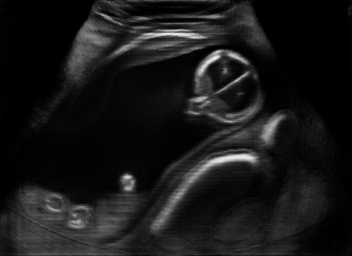}
\includegraphics[width=0.19\textwidth]{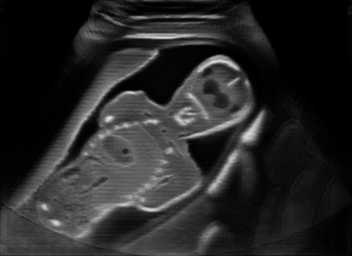}
\includegraphics[width=0.19\textwidth]{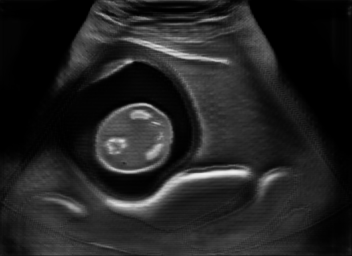}
\includegraphics[width=0.19\textwidth]{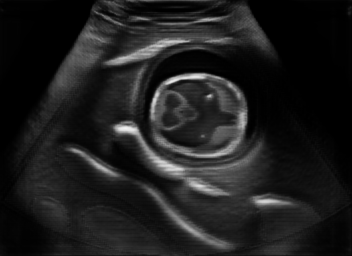}
\includegraphics[width=0.19\textwidth]{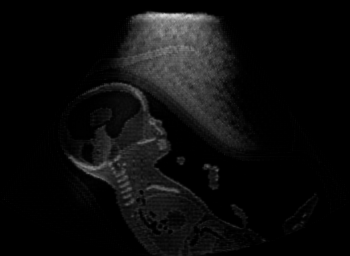}
\includegraphics[width=0.19\textwidth]{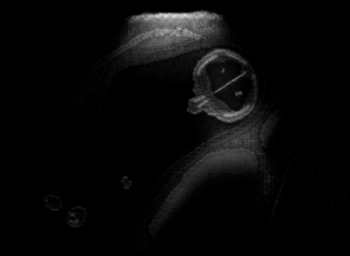}
\includegraphics[width=0.19\textwidth]{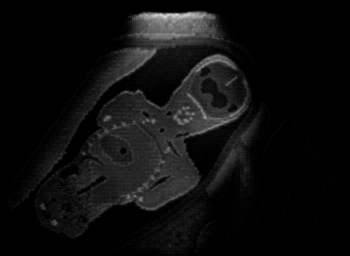}
\includegraphics[width=0.19\textwidth]{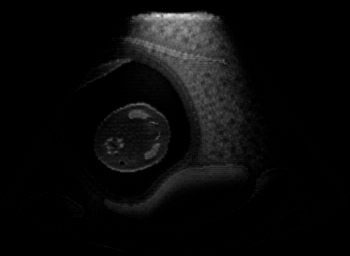}
\includegraphics[width=0.19\textwidth]{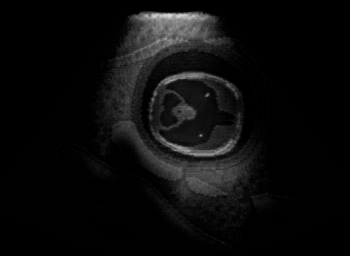}
\caption{Example results of seg-to-real and seg-to-sim image translation: Top: input semantic maps, middle: translated real images, bottom: translated simulation images.}
\label{fig:seg2real}
\end{figure}

\vspace{1ex}\noindent{\bf Discussion. }
Note that, despite both being fetal images, the simulated and the real images have substantially different anatomical contents, which makes the translation task extremely challenging.
Nevertheless, our proposed framework is able to generate images with appearance strikingly close to real images, with far superior realism than its competitors.
Besides sim-to-real translation, given its multi-domain conditional nature, our proposed framework without any further training can also translate images between the other domains, \eg seg-to-real or seg-to-sim, with examples presented in Fig.~\ref{fig:seg2real}.

\section{Conclusions}
We have introduced a contrastive unpaired translation framework with a class-conditional generator, for improving ultrasound simulation realism.
By applying cyclic and semantic consistency constraints, our proposed method can translate domain-specific appearance, while preserving the original content. 
This is shown to outperform state-of-the-art unpaired translation methods.
With the proposed methods, we largely close the appearance gap between simulated and real images.
Future works may include an evaluation of the effects of translated images on US training as well as an investigation of seg-to-real image translation, which can enable to completely dispense with any expensive rendering.

\vspace{1ex}\noindent{\bf Funding} was provided by the Swiss Innovation Agency Innosuisse.

\bibliographystyle{splncs04}
\bibliography{image_translation_us}

\begin{thebibliography}{10}
\providecommand{\url}[1]{\texttt{#1}}
\providecommand{\urlprefix}{URL }
\providecommand{\doi}[1]{https://doi.org/#1}

\bibitem{almahairi2018augmented}
Almahairi, A., Rajeshwar, S., Sordoni, A., Bachman, P., Courville, A.:
  Augmented cyclegan: Learning many-to-many mappings from unpaired data. In:
  International Conference on Machine Learning (ICML). pp. 195--204 (2018)

\bibitem{bargsten2020specklegan}
Bargsten, L., Schlaefer, A.: {S}peckle{G}an: a generative adversarial network
  with an adaptive speckle layer to augment limited training data for
  ultrasound image processing. Int J Comp Asst Radiol Surg (IJCARS)
  \textbf{15}(9),  1427--1436 (2020)

\bibitem{binkowski2018demystifying}
Bi{\'n}kowski, M., Sutherland, D.J., Arbel, M., Gretton, A.: Demystifying mmd
  gans. preprint arXiv:1801.01401  (2018)

\bibitem{burger2013real}
{Burger}, B., {Bettinghausen}, S., {Radle}, M., {Hesser}, J.: Real-time
  {GPU}-based ultrasound simulation using deformable mesh models. IEEE
  Transactions on Medical Imaging  \textbf{32}(3),  609--618 (2013)

\bibitem{choi2018stargan}
Choi, Y., Choi, M., Kim, M., Ha, J.W., Kim, S., Choo, J.: Stargan: Unified
  generative adversarial networks for multi-domain image-to-image translation.
  In: Proceedings of the IEEE Conference on Computer Vision and Pattern
  Recognition (CVPR). pp. 8789--8797 (2018)

\bibitem{goksel2009b}
Goksel, O., Salcudean, S.E.: {B}-mode ultrasound image simulation in deformable
  {3-D} medium. IEEE Transactions on Medical Imaging  \textbf{28}(11),
  1657--1669 (2009)

\bibitem{goodfellow2014generative}
Goodfellow, I., Pouget-Abadie, J., Mirza, M., Xu, B., Warde-Farley, D., Ozair,
  S., Courville, A., Bengio, Y.: Generative adversarial nets. In: Advances in
  Neural Information Processing Systems (NeurIPS). pp. 2672--2680 (2014)

\bibitem{heusel2017gans}
Heusel, M., Ramsauer, H., Unterthiner, T., Nessler, B., Hochreiter, S.: Gans
  trained by a two time-scale update rule converge to a local nash equilibrium.
  In: Advances in Neural Information Processing Systems (NeurIPS). pp.
  6626--6637 (2017)

\bibitem{isola2017image}
Isola, P., Zhu, J.Y., Zhou, T., Efros, A.A.: Image-to-image translation with
  conditional adversarial networks. In: IEEE Conference on Computer Vision and
  Pattern Recognition (CVPR). pp. 1125--1134 (2017)

\bibitem{jafari2020cardiac}
Jafari, M.H., Girgis, H., Van~Woudenberg, N., Moulson, N., Luong, C., Fung, A.,
  Balthazaar, S., Jue, J., Tsang, M., Nair, P., Gin, K., Rohling, R.,
  Abolmaesumi, P., Tsang, T.: Cardiac point-of-care to cart-based ultrasound
  translation using constrained {C}ycle{GAN}. Int J Comp Asst Radiol Surg
  (IJCARS) pp. 1--10 (2020)

\bibitem{jiao2020self}
Jiao, J., Namburete, A.I., Papageorghiou, A.T., Noble, J.A.: Self-supervised
  ultrasound to {MRI} fetal brain image synthesis. IEEE Transactions on Medical
  Imaging  \textbf{39}(12),  4413--4424 (2020)

\bibitem{kingma2014adam}
Kingma, D.P., Ba, J.: Adam: A method for stochastic optimization. ICLR  (2015)

\bibitem{mattausch2017image}
Mattausch, O., Goksel, O.: Image-based reconstruction of tissue scatterers
  using beam steering for ultrasound simulation. IEEE Transactions on Medical
  Imaging  \textbf{37}(3),  767--780 (2017)

\bibitem{mattausch2018realistic}
Mattausch, O., Makhinya, M., Goksel, O.: Realistic ultrasound simulation of
  complex surface models using interactive {Monte-Carlo} path tracing. In:
  Computer Graphics Forum. vol.~37, pp. 202--213 (2018)

\bibitem{mirza2014conditional}
Mirza, M., Osindero, S.: Conditional generative adversarial nets. preprint
  arXiv:1411.1784  (2014)

\bibitem{obukhov2020torchfidelity}
Obukhov, A., Seitzer, M., Wu, P.W., Zhydenko, S., Kyl, J., Lin, E.Y.J.:
  High-fidelity performance metrics for generative models in pytorch (2020),
  https://github.com/toshas/torch-fidelity, accessed 26 Feb 2021

\bibitem{park2020contrastive}
Park, T., Efros, A.A., Zhang, R., Zhu, J.Y.: Contrastive learning for unpaired
  image-to-image translation. In: European Conference on Computer Vision
  (ECCV). pp. 319--345 (2020)

\bibitem{park2019semantic}
Park, T., Liu, M.Y., Wang, T.C., Zhu, J.Y.: Semantic image synthesis with
  spatially-adaptive normalization. In: Proceedings of the IEEE Conference on
  Computer Vision and Pattern Recognition (CVPR). pp. 2337--2346 (2019)

\bibitem{paszke2019pytorch}
Paszke, A., Gross, S., Massa, F., Lerer, A., Bradbury, J., Chanan, G., Killeen,
  T., Lin, Z., Gimelshein, N., Antiga, L., et~al.: Pytorch: An imperative
  style, high-performance deep learning library. preprint arXiv:1912.01703
  (2019)

\bibitem{starkov2019fusion}
Starkov, R., Tanner, C., Bajka, M., Goksel, O.: Ultrasound simulation with
  animated anatomical models and on-the-fly fusion with real images via
  path-tracing. Computers \& Graphics  \textbf{82},  44--52 (2019)

\bibitem{starkov2019ultrasound}
Starkov, R., Zhang, L., Bajka, M., Tanner, C., Goksel, O.: Ultrasound
  simulation with deformable and patient-specific scatterer maps. International
  Journal of Computer Assisted Radiology and Surgery (IJCARS)  \textbf{14}(9),
  1589--1599 (2019)

\bibitem{tom2018simulating}
Tom, F., Sheet, D.: Simulating patho-realistic ultrasound images using deep
  generative networks with adversarial learning. In: IEEE Int Symp on
  Biomedical Imaging (ISBI). pp. 1174--1177 (2018)

\bibitem{tomar2021self}
Tomar, D., Lortkipanidze, M., Vray, G., Bozorgtabar, B., Thiran, J.P.:
  Self-attentive spatial adaptive normalization for cross-modality domain
  adaptation. IEEE Transactions on Medical Imaging  (2021)

\bibitem{vitale2019improving}
Vitale, S., Orlando, J.I., Iarussi, E., Larrabide, I.: Improving realism in
  patient-specific abdominal ultrasound simulation using {C}ycle{GAN}s.
  International Journal of Computer Assisted Radiology and Surgery (IJCARS) pp.
  1--10 (2019)

\bibitem{zhang2021learning}
Zhang, L., Portenier, T., Goksel, O.: Learning ultrasound rendering from
  cross-sectional model slices for simulated training. preprint
  arXiv:2101.08339  (2021)

\bibitem{zhang2020deep}
Zhang, L., Portenier, T., Paulus, C., Goksel, O.: Deep image translation for
  enhancing simulated ultrasound images. In: MICCAI Workshop on Advanaces in
  Simplifying Medical UltraSound (ASMUS), pp. 85--94 (2020)

\bibitem{zhang2020scat}
Zhang, L., Vishnevskiy, V., Goksel, O.: Deep network for scatterer distribution
  estimation for ultrasound image simulation. IEEE Transactions on Ultrasonics,
  Ferroelectrics, and Frequency Control (TUFFC)  \textbf{67}(12),  2553--2564
  (2020)

\bibitem{zhu2017unpaired}
Zhu, J.Y., Park, T., Isola, P., Efros, A.A.: Unpaired image-to-image
  translation using cycle-consistent adversarial networks. In: IEEE
  International Conference on Computer Vision (CVPR). pp. 2223--2232 (2017)

\bibitem{zhu2017toward}
Zhu, J.Y., Zhang, R., Pathak, D., Darrell, T., Efros, A.A., Wang, O.,
  Shechtman, E.: Toward multimodal image-to-image translation. Advances in
  Neural Information Processing Systems (NeurIPS)  (2017)

\end{thebibliography}

\clearpage
\appendix
\section{Appendix: Network Architecture}
\begin{table}
\centering
\caption{\textbf{Generator Architecture.} IN: instance normalization, K: kernel size of convolution layers, shape $h\times w\times ch$: height$\times$weight$\times$number of channels. We use bilinear interpolation for Upsample layers. Note that we use instance normalization only in the encoder.
\textbf{Discriminator Architecture.} K: kernel size of the convolution layers, shape $h\times w\times ch$: height$\times$weight$\times$number of channels.}
\label{tab:generator_arch}
\resizebox{\textwidth}{!}{%
\begin{tabular}{@{}lccr@{}}
\multicolumn{4}{c}{\textsc{Generator}}\\
\toprule
\rowcolor[HTML]{C0C0C0}
\textbf{Layer Name}                         & \textbf{Input Shape} & \textbf{Output Shape}                                                    & \textbf{Layer Information}                                   \\ \midrule
\multicolumn{4}{c}{\cellcolor[HTML]{EFEFEF}\textbf{Encoder}}                                                                                                              \\ \midrule
Down-sampling                & $h\times w\times 1$ & $h\times w\times 64$                                                  & CONV-(K$7\times7$), IN, ReLU                  \\
                             & $h\times w\times 64$ & $\frac{h}{2}\times \frac{w}{2}\times 128$                            & CONV-(K$4\times4$), IN, ReLU                 \\
                             & $\frac{h}{2} \times \frac{w}{2} \times 128$ & $\frac{h}{4} \times \frac{w}{4} \times 256$       & CONV-(K$4\times4$), IN, ReLU                 \\ \midrule
                             & $\frac{h}{4}\times \frac{w}{4}\times 256$ & $\frac{h}{4}\times \frac{w}{4}\times 256$       & Residual Block: CONV-(K$3\times3$), IN, ReLU \\
                             & $\frac{h}{4}\times \frac{w}{4}\times 256$ & $\frac{h}{4}\times \frac{w}{4}\times 256$       & Residual Block: CONV-(K$3\times3$), IN, ReLU \\
                             & $\frac{h}{4}\times \frac{w}{4}\times 256$ & $\frac{h}{4}\times \frac{w}{4}\times  256$       & Residual Block: CONV-(K$3\times3$), IN, ReLU \\
\multirow{-4}{*}{Bottleneck} & $\frac{h}{4}\times \frac{w}{4}\times 256$ & $\frac{h}{4}\times \frac{w}{4}\times 256$       & Residual Block: CONV-(K$3\times3$), IN, ReLU \\ \midrule
\multicolumn{4}{c}{\cellcolor[HTML]{EFEFEF}\textbf{Decoder}}                                                                                                              \\ \midrule
Bottlenect                   & $\frac{h}{4}\times \frac{w}{4}\times (256 + n_c)$ & $\frac{h}{4}\times \frac{w}{4}\times (256 + n_c)$  & Residual Block: CONV-(K$3\times3$), ReLU     \\
                             & $\frac{h}{4}\times \frac{w}{4}\times (256 + n_c)$ & $\frac{h}{4}\times \frac{w}{4}\times (256 + n_c)$       & Residual Block: CONV-(K$3\times3$), ReLU     \\
                             & $\frac{h}{4}\times \frac{w}{4}\times (256 + n_c)$ & $\frac{h}{4}\times \frac{w}{4}\times (256 + n_c)$       & Residual Block: CONV-(K$3\times3$), ReLU     \\
                             & $\frac{h}{4}\times \frac{w}{4}\times (256 + n_c)$ & $\frac{h}{4}\times \frac{w}{4}\times (256 + n_c)$       & Residual Block: CONV-(K$3\times3$), ReLU     \\ \midrule
Up-sampling                  & $\frac{h}{4}\times \frac{w}{4}\times (256+n_c)$ & $\frac{h}{2}\times \frac{w}{2}\times 128$       & CONV-(K$3\times3$), Upsample,  ReLU          \\
                             & $\frac{h}{2}\times \frac{w}{2}\times 128$ & $h\times w\times 64$                            & CONV-(K$3\times3$), Upsample, ReLU           \\ \midrule
Final                        & $h\times w\times 64$ & $h\times w\times 1$                                                  & CONV-(K$3\times3$), Tanh                       \\ \bottomrule
\end{tabular}%
}
\newline
\vspace*{.6cm}
\newline
\resizebox{\textwidth}{!}{%
\begin{tabular}{@{}lccr@{}}
\multicolumn{4}{c}{\textsc{Discriminator}}\\
\toprule
\rowcolor[HTML]{C0C0C0}
\textbf{Layer Name}                                                                     & \textbf{Input Shape} & \textbf{Output Shape}                                                   & \textbf{Layer Information}                                       \\ \midrule
Input Layer                                                               & $h\times w\times 1$ & $\frac{h}{2}\times \frac{w}{2}\times 64$                            & CONV-(K$4\times4$),  Leaky ReLU                   \\ \midrule
\multirow{3}{*}{Hidden Layers}                                            & $\frac{h}{2}\times \frac{w}{2}\times 64$ & $\frac{h}{4}\times \frac{w}{4}\times 128$       & CONV-(K$4\times4$), Leaky ReLU                  \\
                                                                          & $\frac{h}{4}\times \frac{w}{4}\times 128$ & $\frac{h}{8}\times \frac{w}{8}\times 256$      & CONV-(K$4\times4$), Leaky ReLU                  \\
                                                                          & $\frac{h}{8}\times \frac{w}{8}\times 256$ & $\frac{h}{16}\times \frac{w}{16}\times 512$    & CONV-(K$4\times4$), Leaky ReLU                       \\ \midrule
Output Layer: $D_{fake/real}$ & $\frac{h}{16}\times \frac{w}{16}\times 512$ & $\frac{h}{16}\times \frac{w}{16}\times 1$   & CONV-(K$3\times3$)                                 \\
Output Layer: $D_{classifier}$ & $\frac{h}{16}\times \frac{w}{16}\times 512$ & $1\times 1\times n_c$                       & CONV-(K$\frac{h}{16}\times \frac{w}{16}$) \\ \bottomrule
\end{tabular}%
}
\end{table}

\end{document}